# Metastable Chimera States in Community-Structured Oscillator Networks

Murray Shanahan

Department of Computing, Imperial College London, 180 Queen's Gate, London SW7 2AZ, UK.

**Abstract**

A system of symmetrically coupled identical oscillators with phase lag is presented, which is capable of generating a large repertoire of transient (metastable) "chimera" states in which synchronisation and desynchronisation co-exist. The oscillators are organised into communities, such that each oscillator is connected to all its peers in the same community and to a subset of the oscillators in other communities. Measures are introduced for quantifying metastability, the prevalence of chimera states, and the variety of such states a system generates. By simulation, it is shown that each of these measures is maximised when the phase lag of the model is close, but not equal, to $\pi/2$. The relevance of the model to a number of fields is briefly discussed, with particular emphasis on brain dynamics.

Many complex systems, both natural and artificial, exhibit synchronisation phenomena, which can be modeled using weakly coupled oscillators. Most previous synchronisation studies focus on stability. Yet many complex systems (such as the human brain) do not converge on stable synchronised states. Rather they are metastable, temporarily dwelling in the vicinity of one stable state before spontaneously migrating away from it towards another. A second feature of many complex systems (including the brain) is competition. In the context of synchronisation, this is manifest in so-called chimera states, where one coalition of oscillators synchronises while rival coalitions of identical oscillators are desynchronised. This short paper presents the first model to exhibit both chimera states and metastability. This is achieved by organising the oscillators into a community structured (or modular) network (echoing established brain connectivity findings). Measures are introduced to quantify the prevalence of chimera states and metastability. These form the basis of an empirical study that establishes the conditions in which metastability and chimera states are most prevalent. The same study shows (using a novel measure) that the repertoire of metastable states produced under these conditions is also maximised.

## I. INTRODUCTION

Periodic phenomena involving the synchronisation of multiple variables are prevalent both in nature and the human environment, and can be modeled mathematically as systems of coupled oscillators [Pikovsky, *et al*., 2001]. Among the rich variety of behaviours such systems exhibit are states in which a set of identical symmetrically coupled oscillators spontaneously partitions into one subset that is synchronised and another subset that is desynchronised [Kuramoto & Battogtokh 2002; Abrams & Strogatz, 2004; Abrams, *et al*., 2008]. A system that gives rise to these so-called *chimera states* is a plausible model for a competitive process wherein a set of winners forms an alliance to the exclusion of the rest of the population. Since competitive processes of this sort dominate the dynamics of the brain, the economy, and the ecosphere, they are of considerable scientific interest.

Typical studies of synchronisation in systems of coupled oscillators attempt to map their various dynamical regimes and to pin down the conditions for entering those regimes [Acebrón, *et al*., 2005]. Stable states, in which some or all of the oscillators are fully synchronised, have attracted particular attention. However, in many complex systems, extended periods of synchronisation are pathological. Prolonged synchronisation in the brain, for example, is a symptom of seizure [Arthuis, *et al*., 2009]. This motivates the study of *metastability* in systems of coupled oscillators [Niebur, *et al*., 1991; Bressler & Kelso, 2001; Pluchino & Rapisarda, 2006; Kitzbichler, *et al*., 2009]. A system of oscillators exhibits metastability if some or all of its members linger in the vicinity of a synchronised state without falling into such a state permanently. Moreover, in a complex milieu such as the brain, the economy, or the ecosphere, we should expect to witness a *large number* of distinct metastable states.

The upshot of these considerations is that an adequate model of competitive periodic phenomena in systems that are not frozen, static, or in seizure, should exhibit a non-trivial repertoire of metastable chimera-like states. Although the topic has been relatively neglected, metastability in oscillator networks has been described in the literature before. For example, Niebur, *et al*. [1991] reported metastability in a large network of weakly coupled oscillators. But their model included a noise term, and metastability is promoted by the resulting thermal fluctuations [Kuramoto, 1984, Ch. 5]. More recently, Kitzbichler, *et al*. [2009], following the work of Kuramoto [1984], characterised the metastability of a network of oscillators with critical coupling strength. But the natural frequencies of the oscillators in their model are distributed while models of chimera states deploy identical oscillators.

Reinforcing the well-established view that there is a crucial relationship between connectivity and dynamics [Strogatz, 2001; Arenas, *et al*., 2008; Müller-Linow, *et al*., 2008], the required properties are exhibited by the present (deterministic) model because its oscillators (all identical) are arranged in a network with *community structure* [Girvan & Newman, 2002]. That is to say, the nodes of the network (the individual oscillators) are partitioned into subsets (communities or modules) whose members are more densely connected with each other than with nodes outside their community. Since both the functional and structural connectivity of the human brain are similarly modular [Hagmann, *et al*., 2008], the system of oscillators presented here is a plausible model of metastable neural synchronisation.

## II. METHODS AND MEASURES

The present model comprises eight communities of 32 phase-lagged Kuramoto oscillators with identical natural frequencies [Kuramoto, 1984; Acebrón, *et al*., 2005]. Each oscillator is fully connected to its own community, and has 32 random connections to oscillators in other communities. Following the model of Abrams, *et al*. [2008], the intra-community coupling strength is slightly higher than the inter-community coupling strength. All connections are symmetrical.

The phase $\theta_i$ of each oscillator $i$ is governed by the equation

$$\frac{d\theta_i}{dt} = \omega + \frac{1}{N+1} \sum_{j=1}^{N} K_{i,j} \sin(\theta_j - \theta_i - \alpha)$$

where $\omega$ is the natural frequency of the oscillator, $N$ is the total number of connections per oscillator, $K_{i,j}$ is the coupling strength between oscillators $i$ and $j$, and $\alpha$ is a fixed phase lag. In the present model, $\omega=1$, $N=63$, $K_{i,j} = u$ if $i$ and $j$ belong to the same community, $K_{i,j} = v$ if $i$ and $j$ belong to different communities and are connected, and $K_{i,j} = 0$ otherwise. Like Abrams, *et al*. [2008], we define two parameters $A$ and $\beta$ for the model, such that $A = u - v$ where $u + v = 1$, and $\beta = \pi/2 - \alpha$. For the experiment described here $A$ was set at 0.2.

The level of synchrony within a community $c$ at time $t$ may be quantified according to the measure

$$\phi_c(t) = \left| \left\langle e^{i\theta_k(t)} \right\rangle_{k \in c} \right|$$

where $\theta_k(t)$ is the phase of oscillator $k$ at time $t$ and $\langle f \rangle_{k \in c}$ denotes the average of $f$ over all $k$ in $c$. This measure ranges from 0 to 1, where 0 is total desynchronisation and 1 is full synchronisation. (Note that $\phi_c(t)$ quantifies instantaneous synchrony, and does not provide information about coherence.)

By sampling $\phi_c(t)$ for all the communities at discrete intervals, it is possible to quantify both the level of metastability in the system and the prevalence of chimera-like states (Fig. 1). Let $C$ be the set of all $M$ communities, and assume $\phi_c(t)$ is sampled at times $t \in \{1...T\}$ for each $c \in C$. If we fix the community $c$ and estimate the variance $\sigma_{\text{met}}(c)$ of $\phi_c(t)$ over all time points $t \in \{1...T\}$, we get an indication of how much the synchrony in $c$ varies over time. The average of this variance estimate over the set $C$ of all communities is an index of the metastability of the overall system (denoted $\lambda$). So we have

$$\lambda = \langle \sigma_{\text{met}} \rangle_C$$

where

$$\sigma_{\text{met}}(c) = \frac{1}{T-1} \sum_{t \leq T} (\phi_c(t) - \langle \phi_c \rangle_T)^2.$$

Conversely, if we fix the time $t$ and estimate the variance $\sigma_{\text{chi}}(t)$ of $\phi_c(t)$ over all communities in $C$, we get an instantaneous indication of how chimera-like the system is at time $t$. The average of this variance estimate is an index of how chimera-like a typical state of the system is (denoted $\chi$). So we have

$$\chi = \langle \sigma_{\text{chi}} \rangle_T$$

where

$$\sigma_{\text{chi}}(t) = \frac{1}{M-1} \sum_{c \in C} (\phi_c(t) - \langle \phi(t) \rangle_C)^2.$$

Note that if a population $c$ of oscllators is either completely synchronised or completely desynchronised then $\sigma_{\text{met}}(c) = 0$. If $c$ spends equal time in all stages of synchronisation then it presents a uniformly distributed $\phi_c$ with $\sigma_{\text{met}}(c) = \frac{1}{12} \approx 0.083$. A value of $\sigma_{\text{met}}(c)$ greater than $\frac{1}{12}$ is possible if $\phi_c$ has a multi-modal distribution with high peaks at both shoulders. But assuming such distributions do not arise, and as long as the population size is large, we may regard a uniform distribution as indicative of maximum metastability, yielding $\lambda_{\max} = \frac{1}{12} \approx 0.083$. (The population size must be large, since, for example, a pair of decoupled (free-running) oscillators with different frequencies is not metastable yet has uniformly distributed $\phi_c$. By contrast, episodes of high synchrony are rare in a large population of decoupled oscillators with different frequencies, ensuring a non-uniformly distributed $\phi_c$. In the present model, the community size is large, and of course the oscillators are not decoupled.)

Similar considerations apply to $\sigma_{\text{chi}}$. If at some time $t$ all of the communities in the system are either fully synchronised or fully desynchronised then $\sigma_{\text{chi}}(t) = 0$. A "perfect" chimera state might be characterised as one in which exactly half the communities are fully synchronised and half are fully desynchronised (as in the model of Abrams, *et al.* (2008)), which will yield $\sigma_{\text{chi}}(t) = \frac{2}{7} \approx 0.2857$. Although high values of $\sigma_{\text{chi}}$ can be obtained in the present model, due to its metastability they are only transient. So because $\chi$ averages $\sigma_{\text{chi}}$ over time, we should not expect it to approach this maximum. Instead, a metastable system can be considered to attain maximum $\chi$ if it spends half its time in a maximally chimera-like state and half its time in a minimally chimera-like state, which yields $\chi_{\max} = \frac{1}{7} \approx 0.1429$.

In addition to assessing synchronisation within a community, it is possible to quantify pairwise synchronisation across communities. In particular, for every pair of communities $a$ and $b$, we can consider

$$\psi_{a,b}(t) = \left| \tfrac{1}{2} \left( \left\langle e^{i\theta_k(t)} \right\rangle_{k \in a} + \left\langle e^{i\theta_k(t)} \right\rangle_{k \in b} \right) \right|.$$

Note that $\psi_{a,b}(t)$, which ranges from 0 to 1, will only be high if $\phi_a(t)$ and $\phi_b(t)$ are individually high *and* communities $a$ and $b$ are synchronised with each other. If $a$ and $b$ are fully synchronised internally but 180° out of phase with each other then $\psi_{a,b}(t) = 0$.

Although $\lambda$ and $\chi$ taken together can detect the occurrence of metastable chimera states, neither measure can distinguish a system that repeatedly visits the same metastable chimera state from a system that has a large repertoire of metastable chimera states. One way to quantify this repertoire is to assess how "mixed up" is the set of coalitions a system produces over a period of time. Accordingly, we define the (normalised) *coalition entropy* $H_C$ of a system by the equation

$$H_C = -\frac{1}{\log_2 |S|} \sum_{s \in S} p(s) \log_2(p(s))$$

where $S$ is the set of distinct coalitions the system can generate and $p(s)$ is the probability of coalition $s$ arising in any given time point. The measure is normalised to lie to between 0 and 1. In the context of the present model we can consider coalitions of synchronised communities. A coalition $s$ is said to arise at time $t$ if $\phi_c(t) > \gamma$ for all $c \in s$, where $\gamma$ is a synchronisation threshold (we shall use $\gamma = 0.8$). Clearly for a system of $M$ communities there are $2^M$ possible coalitions, so $\log_2 |S| = M$. If all $2^M$ possible coalitions arise with equal probability, we have $H_C = 1$. On the other hand, if the system resides permanently in the same state (whether fully synchronised, fully desynchronised, or any sort of chimera-like state), then only one coalition arises and we have $H_C = 0$.

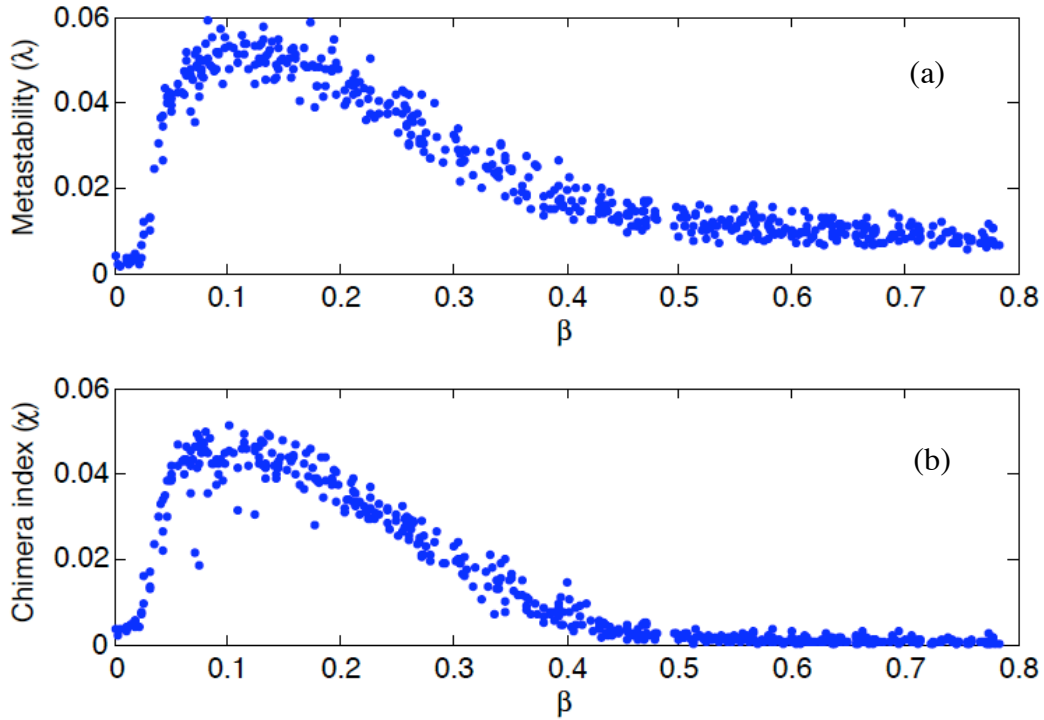

FIG. 1: The behaviour of the model for randomly generated values of $\beta$ between 0 and $\pi/4$ over 500 trials. Initial phases were randomised for each trial. Both metastability index ($\lambda$) and chimera index ($\chi$) are close to zero for $\beta = 0$, peak at around $\beta = 0.1$, and tail off rapidly.

## III. RESULTS

A series of 1000-step trials of for a range of values of $\beta$ was carried out. 500 trials were performed for randomly generated values of $\beta$ ranging from 0 to $\pi/4$. All numerical simulation was carried out using the 4$^{\text{th}}$-order Runge-Kutta method with a step size of 0.05. Intra-community coupling assignments and initial phases were randomised for each trial. The internal synchrony $\phi_c(t)$ was calculated for each community at 5-step intervals, and the resulting data was used to compute $\lambda$, $\chi$ and $H_C$. In addition, an index of global synchrony $\Psi$ was calculated, taken as the average of $\phi_c(t)$ over all times and communities.

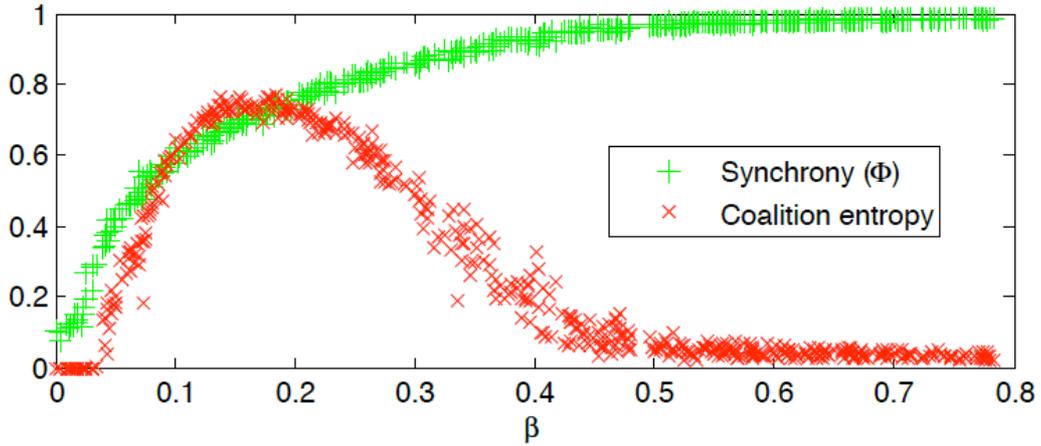

FIG. 2: Global synchrony ($\Phi$) and coalition entropy for the set of trials depicted in Figure 1. Coalition entropy presents a similar profile to metatsability and chimera index (Fig. 1), although it peaks slightly later, at around $\beta = 0.15$.

The results are presented in scatter plots of Figs. 1 & 2. These figures suggest that the model behaves as advertised, and is capable of generating a large repertoire of metastable chimera-like states, but only when $\beta$ falls within a certain narrow range. Metastability and chimera indices are maximised when $0.05 < \beta < 0.15$ (Fig. 1), at which point $0.6 < \Psi < 0.7$ (Fig. 2). When $\beta = 0$ (ie: the phase lag is exactly $\pi/2$), the system finds it hard to synchronise at all, and each of the measures is correspondingly low. At the other end of the scale, when $\beta > \pi/8$, the system tends towards full synchronisation in all communities, and metastability, chimera index, and coalition entropy all tail off accordingly. Coalition entropy peaks slightly later than the other two measures, with $0.1 < \beta < 0.2$ (Fig. 2).

As with other systems of oscillators that exhibit metastability, fluctuations of synchronisation and desynchronisation are prevalent only in a narrow, critical region that resembles a thermodynamic phase transition from order to disorder [Kuramoto, 1984, Ch. 5]. The area to the left of the critical region is the disordered regime, wherein the phases of the oscillators are subject to predominantly repelling forces, while the area to the right of the critical region is the ordered regime, wherein attracting forces dominate. In the critical region there is a balance of repelling and attracting forces. But this balance is not static, and the dominant force applicable to each community alternates between phase attraction and phase repulsion. Since the oscillators have identical natural frequencies and there is no external stochastic perturbation, the reasons for this are not clear, and further work is required to understand the mechanisms underlying the model's behaviour.

To gain some preliminary insight into the behaviour of the model in the critical region, we shall examine a single run in detail. Figure 3 (a) shows the evolution of synchrony ($\phi$) within all eight oscillator communities over a 200 time-step interval in a typical trial with $\beta \approx 0.1$. The trial is representative of the dynamical regime of most interest to us here. The statistics for this trial are: $\chi = 0.0525$, $\lambda = 0.0542$, $H_C = 0.6341$, and $\Phi = 0.5798$. The system exhibits several chimera-like states in which some oscillator communities are highly synchronised while others are desynchronised. From time 250 to 270, for example three communities are highly synchronised ($\phi > 0.8$) and three are desynchronised ($\phi < 0.4$), while two communites have intermediate values of $\phi$. A similarly clear chimera-like state can be seen from time 310 to 330, but with a different combination of synchronised communities.

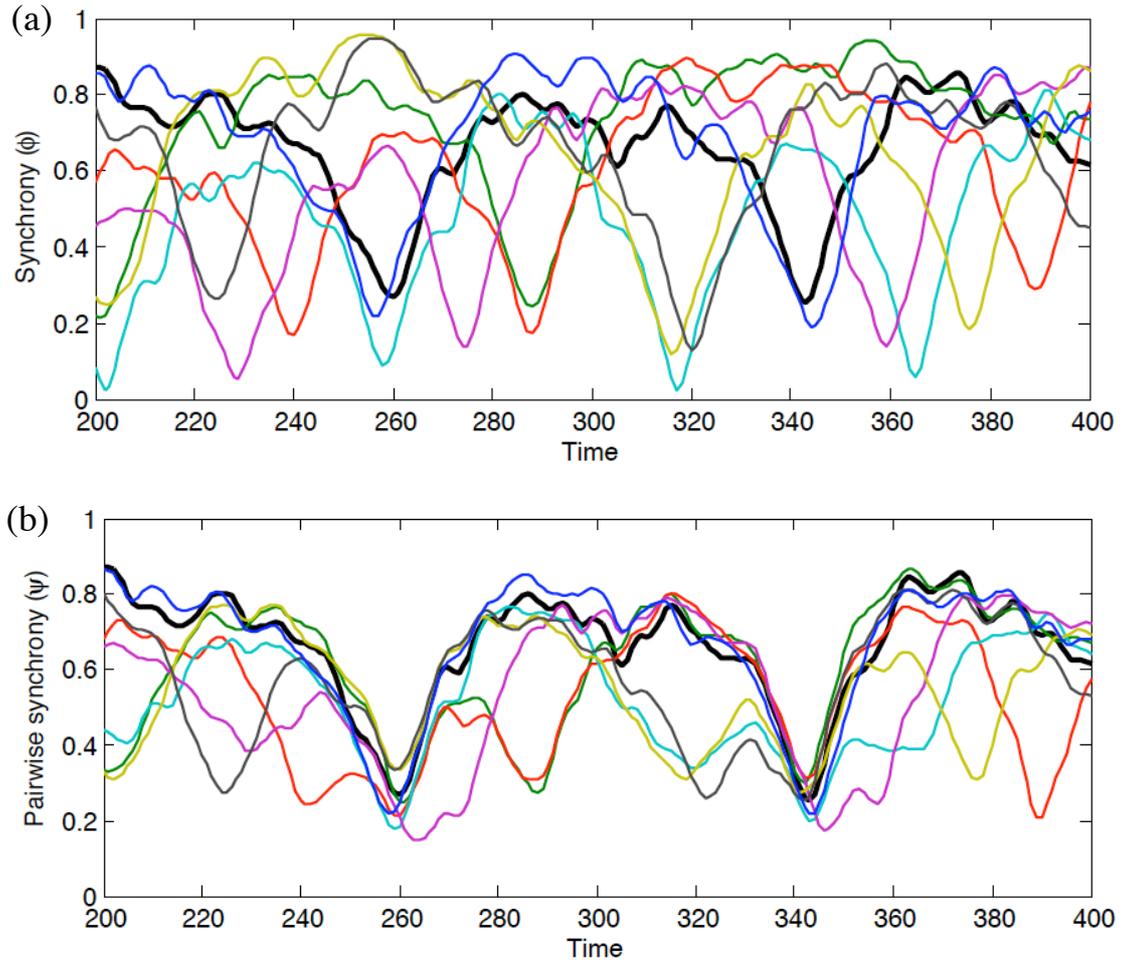

FIG. 3: A 200 time-step period in a typical run with $\beta \approx 0.1$. (a) Intra-community synchrony: The system exhibits several clear chimera-like states. From time 250 to 270, for example three communities are highly synchronised and three are desynchronised. (b) Inter-community synchrony: Pairwise synchrony is plotted between one selected community (shown in black) and each of the eight communities (including itself). From time 310 to 330 the selected community is synchronised with several others, forming a temporary coalition.

Fig. 4 shows how synchrony is distributed over time for each of the eight communities in the same sample run. As expected from Fig. 3 (a), none of the communities spends the majority of its time in any one stage of internal synchronisation. Despite a tendency towards synchrony (a rightwards skew), the resulting distributions all have high variance. Its noteworthy, however, that different communities exhibit different profiles, with Community 2, for example, showing a more pronounced tendency to internally synchronise than its peers.

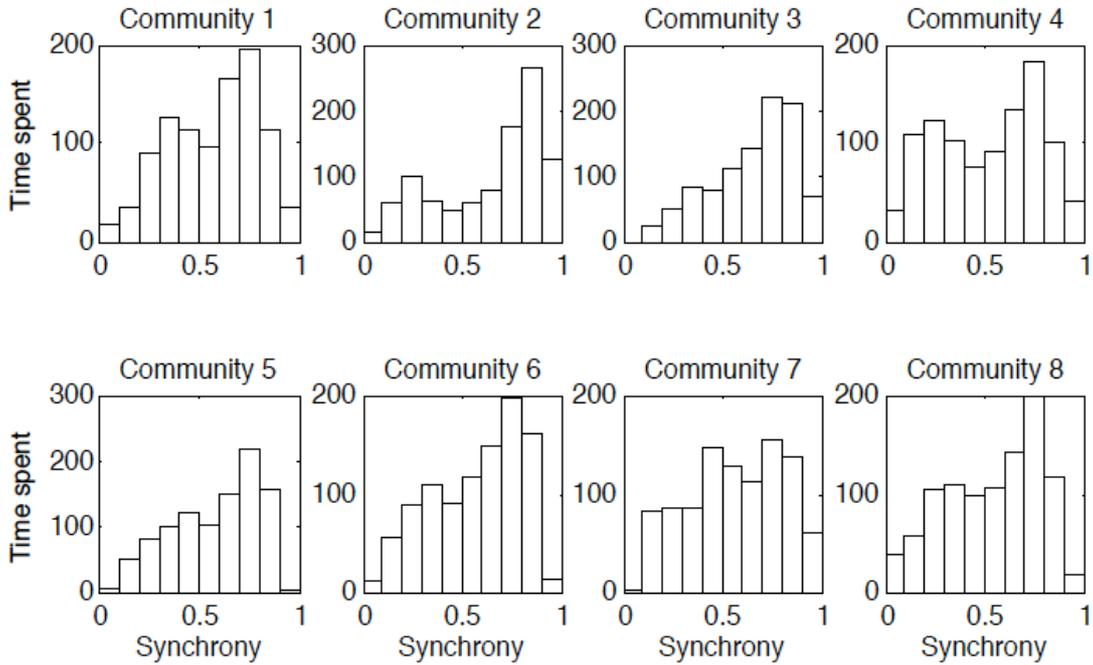

FIG. 4: The distribution of $\phi$ for each community in the run depicted in Figure 3. The distributions all have variances far from zero, indicating metastability in the sense that the oscillator communities spend time in all stages of synchronisation.

The plots in Fig. 3 (a) are highly irregular, and there is no clear pattern to the coincidence of peaks and troughs, which suggests that the system is capable of generating a large repertoire of coalitions of simultaneously synchronised communities. By contrast, the equivalent plot for a "breathing" chimera state has a regular rhythm (see Fig. 2 of [Abrams, *et al*., 2008]). The lack of any discernible regularity in the sequence of metastable states visited by the system hints at both chaotic itinerancy [Kaneko & Tsuda, 2003; Tsuda, *et al*., 2004] and dynamical complexity [Tononi, *et al*., 1998; Seth, *et al*., 2006; Shanahan, 2008b], but further work would be required to establish these properties rigorously.

Further insight into the generation of coalitions by examining the ebb and flow of pairwise synchronisation ($\psi$). Fig 3 (b) shows, for the same 200-step period of the same trial, the pairwise synchrony $\psi_{a,b}$ between a selected community $a$ and all communities $b$ (including $a$ itself). As noted in the previous section, the pairwise synchronisation between two oscillator communities will be low if either community has low internal synchronisation. In the troughs near times 260 and 340, all pairwise synchrony measures are low because the internal synchrony of community $a$ is itself low. At other times, the internal synchrony of $a$ is high, as is the pairwise synchrony between $a$ and various peers. Sets of such synchronised peers constitute temporary coalitions, whose constitution varies over time. At time 290, for example, $a$ is in a coalition alongside five of its peers, with two communities excluded. But by time 320 three communities have been expelled from the coalition, while the two previous exclusions have been recruited into its membership.

## IV. DISCUSSION

Although the essential characteristic of the model — the ability to generate a large repertoire of metastable chimera states — reflects properties common to many real-world complex dynamical systems, the task remains of mapping each of those systems onto the model. For example, it has been proposed that synchronised oscillations in the brain permit effective co-operation among distinct populations of neurons [Fries, 2005; 2009; Womelsdorf, *et al*., 2007], while phenomena such as binocular rivalry, inattentional blindness, and the Stroop effect attest to their competitive character. In other words, the dynamics of the brain seems to arise from the interplay of co-operation and competition, resulting in the formation of synchronised coalitions [Doesburg, *et al*., 2009]. Moreover, thanks to an animal's ongoing activity, its brain is endlessly subject to an open-ended variety of perturbations, and to respond effectively these coalitions must be in constant flux. The dynamics of the present model exhibits the same combination of features. However, to date there is no neurologically detailed model to match, although several existing spiking neuron models deal with relevant issues, such as cortical competition [Dehaene, *et al*., 2003; Deco & Rolls, 2005; Shanahan, 2008a], community structure [Shanahan, 2008b], and the interplay of synchronisation and desynchronisation [Tsuda, *et al*., 2004].

The findings reported here raise a number of questions to be addressed in future work. The issues of chaotic itinerancy and dynamical complexity have already been mentioned. Another urgent task is to develop a theoretical understanding of the phenomena described. The obvious starting point is the analytical treatment of Abrams, *et al*. [2008], whose model is the basis for the present work. However, analogous results may be difficult to obtain for the present model, given its more complex network structure and the variety of synchronisation effects it displays. Finally, it would be fruitful to investigate the occurrence of metastable chimera states in networks with *hierarchical* community structure (especially as brain networks exhibit this property [Zhou, *et al*., 2006; Ferrarini, *et al*, 2009]), and to attempt to characterise the "path" to such states in the manner of Arenas, *et al*. [2006].